# Vibrations induites dans les sols par le trafic ferroviaire : expérimentations, modélisations et isolation
*Railway vibrations induced into the soil: experiments, modelling and isolation.*

**Jean-François Semblat (LCPC), Luca Lenti (LCPC), Delphine Jacqueline (CER Rouen), Jean-Jacques Leblond (LR Clermont), Eva Grasso (LCPC/LMS)**

LCPC, Département Géotechnique, Eau et Environnement, Groupe Séismes et Vibrations, 58 bd Lefebvre, 75732 Paris Cedex 15.
Centre d'expérimentation routière, CETE Normandie-Centre, 10, Chemin de la Poudrière, B.P. 245, 76121 Grand-Quevilly Cedex.
LR Clermont-Ferrand, CETE de Lyon, 8-10, rue Bernard Palissy ZI du Brézet, BP 11, 63014 Clermont-Ferrand Cedex.
Laboratoire de Mécanique des Solides, Ecole Polytechnique, 91128 Palaiseau Cedex.

**Résumé**

Le trafic ferroviaire induit des sollicitations cycliques et dynamiques dans la structure de la voie mais également dans le sol support et l'environnement (Degrande et al. 2006, François et al. 2007, Kausel 2008, Lefeuve-Mesgouez et al 2002, Paolucci et Spinelli 2006). L'analyse de ces sollicitations et des effets induits (e.g. vibrations, ondes…) est fondamentale pour apprécier leur ampleur et remédier à leurs conséquences éventuelles (tassements, nuisances…).
Après un bref rappel de la réglementation, des expérimentations in situ montrent tout d'abord la variabilité des paramètres caractérisant les principaux phénomènes en jeu (propagation d'ondes dans le sol, vibrations induites…). Les principaux essais dynamiques en laboratoire sont ensuite présentés. Ils autorisent la détermination des caractéristiques dynamiques des matériaux (e.g. essais à la colonne résonnante), mais aussi une analyse simplifiée des phénomènes vibratoires en conditions contrôlées (e.g. essais en fosse géotechnique, essais en centrifugeuse).
Après avoir caractérisé les sources vibratoires et les contrastes de raideur (ou de vitesse d'ondes) entre les différentes couches de sol (ou diverses inclusions), il est alors possible de modéliser des configurations types ou réalistes à l'aide de méthodes théoriques (fonctions de transfert) ou numériques (e.g. : éléments finis, éléments de frontière). Des études paramétriques permettent d'analyser les phénomènes de propagation et l'amortissement dans le sol afin d'estimer l'évolution spatiale de l'amplitude des vibrations dans ces différentes configurations.
In fine, il peut être nécessaire d'envisager, le cas échéant, des techniques de mitigation ou d'isolation afin de limiter les conséquences éventuelles des vibrations induites. Plusieurs résultats expérimentaux et numériques originaux illustreront ce dernier point.

**Abstract**

Railway traffic induces cyclic and dynamic loadings in the track structure but also in the close environment (Degrande et al. 2006, François et al. 2007, Kausel 2008, Lefeuve-Mesgouez et al 2002, Paolucci et Spinelli 2006). The analysis of such excitations and their effects (e.g. vibrations, waves, etc) is fundamental to estimate their level and mitigate their potential consequences (settlements, nuisances, etc).
After a brief summary of the current regulations, in situ experiments show the variability of the parameters characterizing the main phenomena (wave propagation into the soil, induced vibrations, etc). The main dynamic laboratory experiments are then discussed. They allow the estimation of the dynamic features of the materials (e.g. resonant column test), but also a simplified analysis of the main phenomena under controlled conditions (e.g. experiments in a geotechnical pit, centrifuge tests).

The vibratory sources and the impedance ratios between the various soil layers (or some inclusions) being known, it is then possible to model some specific or actual configurations through theoretical (transfer functions) or numerical (e.g. finite elements, boundary elements) methods. Parametric studies allow the analysis of the propagation phenomena and the attenuation process in the soil in order to investigate the spatial variations of the vibrations amplitude in such various configurations.

Finally, it may be useful to consider mitigation or isolation techniques in order to limit the consequences of the induced vibrations (e.g. vibratory nuisances, radiated noise). Several experimental and numerical results illustrate this key issue.

## 1 Introduction

Dans cet article, différents aspects des problèmes vibratoires sont abordés. Un bilan des aspects réglementaires est d'abord proposé. En partant de données mesurées à proximité de voies ferroviaires, on constate ensuite la variabilité très forte des vitesses particulaires maximales en fonction du type de train mais aussi, et surtout, des caractéristiques des sols. La caractérisation dynamique des sols en laboratoire (colonne résonante, essais en centrifugeuse) est donc abordée dans la suite de l'article. Les paramètres prépondérants sont le module dynamique (ou la célérité des ondes) et l'amortissement. Comme la propagation des ondes et vibrations dépend fortement des constrastes de propriétés entre les couches, quelques éléments théoriques en attestant sont ensuite rappelés. Pour les applications réalistes, des simulations numériques utilisant les paramètres dynamiques déterminés sur site ou en laboratoire sont nécessaires (méthode des éléments finis (FEM), méthode des éléments de frontière (BEM)). La modélisation est un outil indispensable pour analyser les mesures sur site et généraliser les lois d'atténuation donnant les vitesses particulaires maximales en fonction de la distance pour différents types de sols. Elles permettent également des comparaisons avec les expérimentations en conditions contrôlées pour différentes sources d'excitation (e.g. essais en fosse géotechnique) et sont particulièrement adaptées à l'analyse et l'optimisation de dispositifs d'isolation vibratoire. Ce dernier point est abordé en dernière partie d'article à l'aide de mesures sur site originales et d'exemples de simulations récents.

## 2 Aspects réglementaires

Il n'existe, à ce jour, aucune réglementation spécifique aux vibrations ferroviaires. Seuls s'appliquent les textes généraux sur l'environnement et le droit des tiers imposant la prise en compte des nuisances vibratoires, sans indication de valeur limite ni de méthode d'évaluation des effets. Cette situation devrait toutefois évoluer, les autorités compétentes ayant engagé une réflexion sur l'opportunité de modifier le cadre normatif, réglementaire et législatif dans le domaine des vibrations. La normalisation connaît en revanche une évolution récente importante émanant de trois approches différentes des phénomènes vibratoires exposées ci-après.

La norme NF ISO 14837-1 « Vibrations et bruits initiés au sol dus à des lignes ferroviaires - partie 1 - directives générales » (avril 2006), a été élaborée par la commission ISO/TC108 homologuée NF, et représente le seul texte spécifiquement consacré aux vibrations ferroviaires. Elle décrit successivement, les phénomènes vibratoires concernés, leurs effets et leur mode de mesurage et d'évaluation. Elle définit ensuite différents types de modèles prédictifs suivant une complexité croissante avec l'état d'avancement des études de l'infrastructure concernée. Ce texte doit être complété à court terme par cinq autres parties concernant les modèles prédictifs, le mesurage, les critères d'évaluation, les mesures d'atténuation et la gestion des actifs.

La norme française NF E 90-020 « Méthode de mesurage et d'évaluation des réponses des constructions, des matériels sensibles et des occupants » (juillet 2007), s'applique à toutes les vibrations provoquées par l'activité humaine, dont les vibrations ferroviaires. Ce texte définit trois modes de mesurage et d'évaluation des vibrations en fonction du mode d'intervention (expertise, études, contrôles) et des instrumentations dépendant du récepteur (structure, équipements sensibles, occupants).

Existent également des normes spécifiques dépendant du récepteur comme la norme NF ISO 8569 concernant les équipements sensibles, la norme NF EN ISO 8041 (E90-403) concernant l'appareillage de mesure pour évaluer la réponse des individus aux vibrations. Ces dernières normes définissent des méthodes de mesurage et d'évaluation des effets des vibrations en fonction du récepteur qui les supporte et s'appliquent aux vibrations ferroviaires lorsqu'elles sollicitent ces types de récepteur.

On constate toutefois un manque de cadre législatif global permettant d'estimer l'impact des vibrations sur les structures, les ouvrages et les équipements sensibles, pour quantifier l'inconfort des individus ou pour apprécier les éventuelles responsabilités vis-à-vis des dommages ou des nuisances occasionnés. Cette situation reflète l'importance des enjeux liés à ces thématiques et la complexité des phénomènes physiques sous-jacents. Il s'avère donc nécessaire de faire évoluer les normes d'une part et l'état des connaissances d'autre part. Sur ce dernier point, plusieurs aspects doivent être approfondis : modalités d'émission des vibrations, propagation des ondes dans les structures des voies et dans les sols, effets sur l'environnement urbain et industriel. Pour une source donnée et un même trajet de propagation, les effets des vibrations sont également conditionnés par l'état des structures (notamment leur vulnérabilité) et par la sensibilité des individus. Ces considérations constituent un enjeu majeur pour les normes sur les vibrations ferroviaires puisque celles-ci définissent les paramètres à contrôler et les valeurs seuil à respecter. Une compréhension approfondie des risques et nuisances associés aux vibrations induites par le trafic ferroviaire est donc indispensable.

## 3  Mesures des vibrations sur site et caractérisation des sols en laboratoire

Afin d'étudier les phénomènes liés à l'émission et la propagation d'ondes et de vibrations dues au trafic ferroviaire, il est indispensable de mener des campagnes de mesure sur le terrain. Elles permettent d'appréhender la réalité dans toute sa complexité, mais elles combinent donc simultanément toutes les incertitudes concernant les caractéristiques de la source (interaction roues-voie, structure de la voie ferrée, etc), les différents sols traversés et les hétérogénéités éventuellement présentes sur le trajet de propagation des ondes (cavités, formations géologiques particulières, irrégularités topographiques, etc). Si les expérimentations in situ permettent une évaluation réaliste des principaux phénomènes en jeu, les essais en laboratoire ou en conditions contrôlées autorisent soit la détermination des caractéristiques dynamiques des matériaux, soit une analyse simplifiée des phénomènes (notamment amplification et amortissement).

### 3.1  Mesure des vibrations ferroviaires sur site

Les mesures effectuées aux abords des réseaux ferroviaires existants montrent une extrême variabilité des phénomènes vibratoires due à la multiplicité des paramètres influents provenant du matériel roulant, de la conception et l'état de l'infrastructure et des propriétés des sols environnants.

Il est possible, à partir d'un échantillon de données important, d'évaluer l'influence relative de certains de ces paramètres par analyse statistique de l'évolution d'un indicateur vibratoire isolé. Ce type d'étude, à vocation plus qualitative que quantitative, permet à la fois d'orienter puis de valider les conclusions de programmes de recherche plus élaborés en environnement contrôlé.

Ce type d'étude est illustré sur la Figure 1 qui synthétise les résultats de mesures effectuées en 2006 sur 29 sites de la région Auvergne au passage de 133 trains. Chaque site était instrumenté à partir de trois capteurs de vitesse de vibration implantés sur le sol à des distances variant de 3 à 65m de la voie.

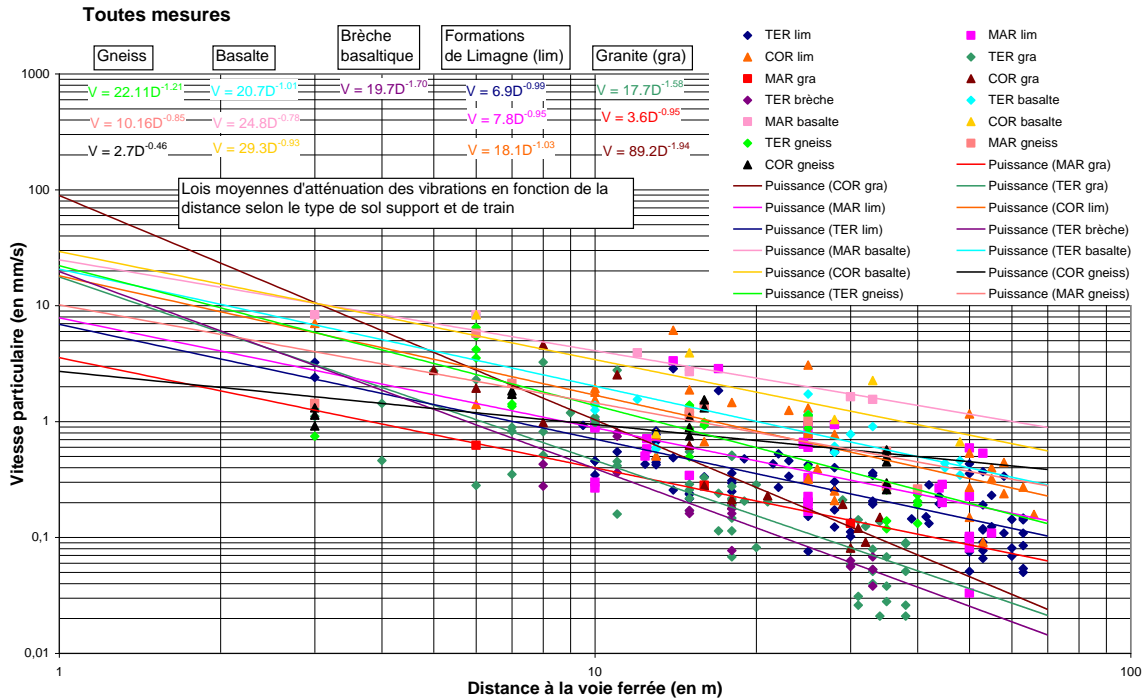

**Figure 1** : Vitesse particulaire mesurée en 2006 sur 29 sites de la région Auvergne au passage de 133 trains différents (source CETE Lyon).
*Particle velocity measured in 2006 for 29 sites at the passage of 133 different trains in the Auvergne region (source CETE Lyon).*

L'indicateur vibratoire utilisé est la valeur de vitesse particulaire maximale enregistrée par point de mesure sur la plage 1-150 Hz. Les paramètres influents analysés sont : la distance à la voie, la nature géologique du sol support, le type de train (TER, corail ou fret).

L'analyse de ce diagramme permet d'évaluer l'influence de ces paramètres sur les vibrations émises et notamment :
1) des vitesses de vibration plus élevées à courte distance de la voie sur les terrains rocheux (granite, gneiss) mais s'atténuant plus fortement avec la distance que sur les terrains meubles (formation de Limagne),
2) des vitesses de vibration nettement plus faibles pour les TER, tous terrains confondus, les trains « grandes lignes » et « fret » générant des vitesses plus élevées,
3) des atténuations des vitesses de vibration en fonction de la distance plus faibles pour les trains de fret.

La complexité des résultats est telle que des analyses fines en laboratoire, en conditions contrôlées ou via des simulations s'avèrent indispensables. De nombreux résultats de mesures in situ sont disponibles dans la littérature (Degrande et al 2006, Paolucci et al 2003).

### 3.2 Caractérisation dynamique des sols en laboratoire

L'essai de colonne résonnante est l'une des méthodes de laboratoire les plus usitées pour la caractérisation dynamique des sols (Cascante et al., 2005; Chung et al., 1984; Drnevich & Richart, 1970; Luong, 1986; Saxena & Reddy, 1989). La qualité de l'essai dépend essentiellement de l'homogénéité des contraintes et des déformations dans l'échantillon. Comme décrit sur la Figure 2, la méthode consiste à solliciter un échantillon de sol en vibration (longitudinale, transversale ou de torsion). Le dispositif expérimental peut être utilisé en sollicitations forcées ou libres (Semblat et Pecker, 2009). Les deux méthodes autorisent l'estimation du module du sol mais également de son amortissement plus difficile à mesurer sur site.

Quand l'essai est réalisé en vibrations forcées, la fréquence doit être ajustée de façon à atteindre la résonance de l'échantillon. La première fréquence propre de l'échantillon ($f_1$) peut alors être estimée d'après le pic de la courbe amplitude/fréquence (Figure 2, droite). La résonance se produisant pour un quart de longueur d'onde, cela permet, dans le cas de la torsion, le calcul de la célérité $V_S$ des ondes de cisaillement dans l'échantillon et donc du module de cisaillement du sol, soit :

$$\mu = \rho V_S^2 = \rho (4 f_1 L)^2 \tag{1}$$

où $f_1$ est la fréquence de résonance de l'échantillon et $L$ est sa hauteur.

L'essai permet également la détermination de l'amortissement du sol. En effet, la courbe de résonance obtenue en vibrations forcées (Figure 2, droite) est caractérisée par une largeur de bande $\Delta f$. L'amortissement en est déduit à partir de l'expression suivante :

$$\xi = \frac{\Delta f}{2 f_1} = \frac{1}{2Q} \tag{2}$$

où $Q$ le facteur de qualité.

Il est également possible de réaliser des essais en vibration libre en arrêtant la sollicitation vibratoire de façon instantanée. L'amortissement du sol peut alors être estimé en analysant la décroissance temporelle d'amplitude définie à l'aide du décrément logarithmique $\delta$ sous la forme (avec $x_i$ amplitude maximale du cycle $i$) :

$$\delta = \ln \frac{x_{n+1}}{x_n} = 2\pi\xi \tag{3}$$

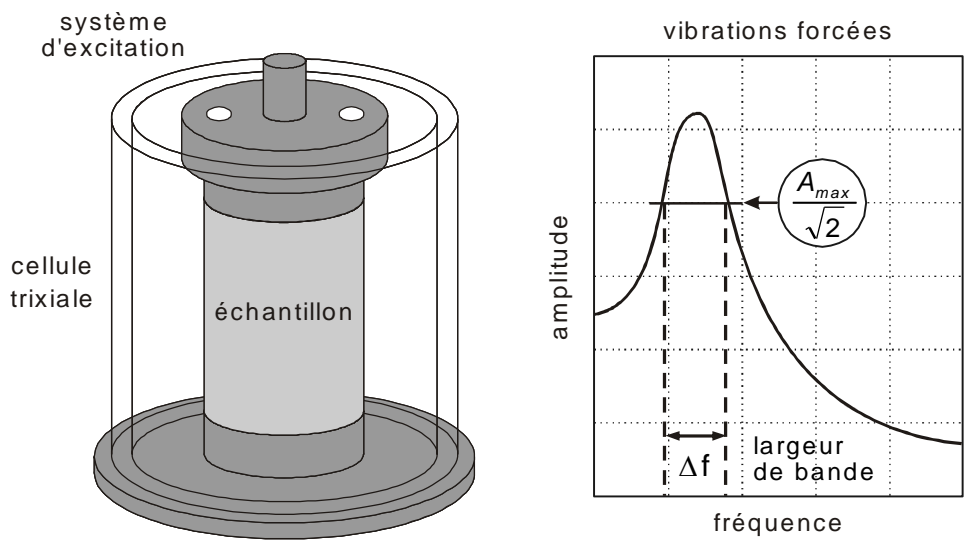

**Figure 2** : Caractérisation dynamique des sols en laboratoire : dispositif de colonne résonnante (gauche) et réponse typique d'un échantillon en vibration forcée (droite).
*Dynamic characterization of soils in the lab: resonant column test (left) and typical specimen response in forced vibrations (right).*

L'essai de colonne résonnante présente la même souplesse d'utilisation qu'un essai triaxial. Il permet de mesurer les caractéristiques des sols pour des amplitudes de déformation comprises entre $10^{-6}$ et $5.10^{-4}$ environ pour les essais en torsion, et pour des amplitudes plus faibles en compression. L'essai de colonne résonnante permet d'obtenir le module maximal qui peut être, dans certaines conditions, directement comparé à celui déduit de mesures géophysiques en place. Pour la détermination de ce module maximal, seule la connaissance de la fréquence de vibration et de la configuration géométrique de l'appareillage est requise. Aucune mesure de déformation n'est théoriquement nécessaire, bien que celle-ci soit effectuée. La précision de la mesure est donc accrue par rapport à un essai où le module est obtenu par mesure de la force appliquée et de la déformation résultante.

### 3.3   Essais dynamiques sur modèles réduits

L'analyse de la propagation d'ondes et de vibrations dans les sols peut être réalisée en conditions contrôlées grâce à des expérimentations sur modèles réduits (Arulanandan et al., 1982, Chazelas et al 2003, Cheney et al., 1990, Coe et al., 1985, Semblat & Luong, 1998). Les expérimentations en macrogravité (centrifugeuse) permettent de reproduire, à échelle réduite, l'état de contrainte observé dans un massif de dimensions réelles. Ces essais permettent ainsi de réaliser des analyses paramétriques difficilement envisageables sur un site réel.

Différents dispositifs expérimentaux sont présentés sur la Figure 3. Les essais sur modèle réduit centrifugé permettent de caractériser finement les ondes et les vibrations créées dans les sols (directivité, propagation, amortissement). Les dimensions habituelles du massif de sol centrifugé sont de l'ordre de 0,5 à 1,5 m, soit l'équivalent de 50 à 150 m pour une pesanteur artificielle de 100 g (Semblat & Pecker, 2009). Quelques exemples de mesures sont donnés dans le paragraphe suivant où ils sont comparés à des analyses théoriques de l'atténuation des ondes dans le sol.

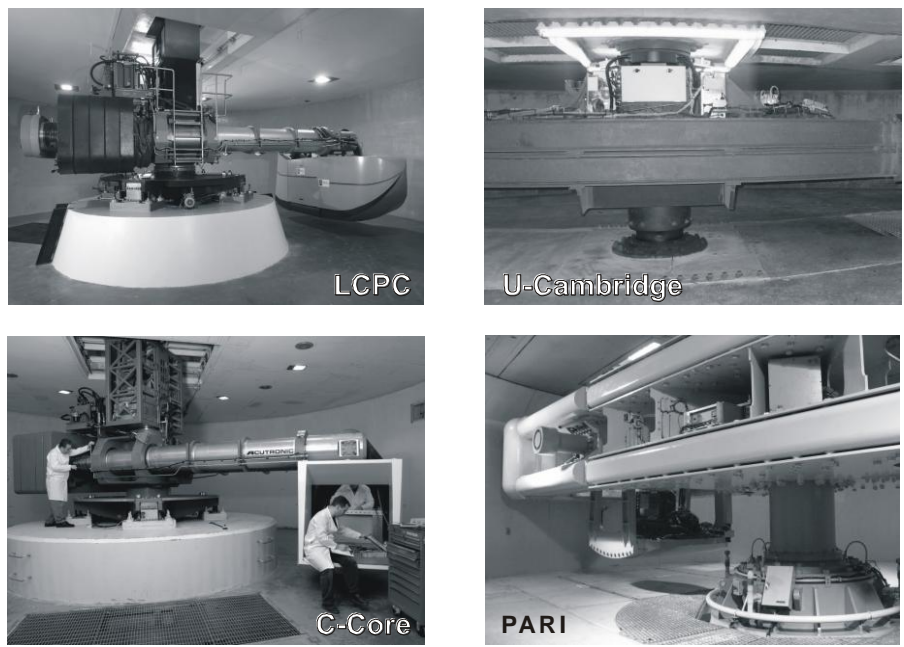

**Figure 3** : Dispositif expérimental pour les essais sur modèles réduits en centrifugeuse : LCPC Nantes, Université de Cambridge, centre C-Core à Terre-Neuve, PARI au Japon (Semblat et Pecker, 2009).
*Experimental facilities for reduced scale centrifuge experiments: LCPC Nantes, University of Cambridge, C-Core center in New Foundland, PARI in Japon (Semblat & Pecker, 2009).*

## 4 Modélisation de la propagation des vibrations

La compréhension des mesures de vibration réalisées sur site est partiellement envisageable à partir d'une analyse mécanique des phénomènes (en terme de valeurs de vitesse, d'accélération, de déplacement, de déformation, de contrainte, etc). La confrontation d'une telle analyse avec les donnés expérimentales permet une première interprétation de celles-ci. Cependant, les données de terrain faisant souvent défaut, seuls les ordres de grandeur sont généralement accessibles. En effet, l'atténuation ou l'amplification du mouvement vibratoire et l'évolution du contenu en fréquence du champ d'onde sur le trajet de propagation dépendent des modules (ou célérités d'ondes) et des amortissements dans les couches et de leurs épaisseurs respectives. En outre, dans les configurations géométriques complexes, les modèles mécaniques échappent à une résolution exacte et il est nécessaire d'envisager des simulations numériques. Les deux approches (mécanique et numérique) sont discutées ci-après.

### 4.1 Modélisation théorique de l'atténuation

A partir de mesures accélérométriques réalisées dans un massif de sol, il est possible d'étudier la propagation et l'amortissement des ondes en utilisant un modèle viscoélastique linéaire. En effet, à partir d'un signal mesuré à la distance $x_i$ de la source, l'accélération à la distance $x_j$ peut être estimée, dans le domaine des fréquences, à l'aide de la relation suivante (Semblat et Pecker, 2009) :

$$a^*(x_j,\omega) = a^*(x_i,\omega) \exp\left[ik^*(\omega)(x_j - x_i)\right] \quad (4)$$

où $a^*(x,\omega)$ représente le spectre de Fourier de l'accélération mesurée à la distance $x$ et $k^*(\omega)$ le nombre d'onde complexe défini par (Aki et Richards 1980) :

$$k^*(\omega) = k(\omega) - i\alpha(\omega) \quad (5)$$

où $k(\omega) = 2\pi c/\omega$ est le nombre d'onde réel (proportionnel à la célérité $c$ de l'onde) et $\alpha(\omega)$ est le facteur d'atténuation déterminé à partir du facteur de qualité $Q(\omega)$ ou du module complexe $M(\omega)$ du modèle viscoélastique considéré (Bourbié et al. 1987).

En utilisant une transformation de Fourier pour obtenir le signal temporel en $x_j$, la comparaison des simulations obtenues d'après cette résolution analytique avec les résultats expérimentaux permet d'identifier les caractéristiques d'amortissement du sol. Suivant le modèle rhéologique retenu, l'atténuation ($Q^{-1}$) varie différemment en fonction de la fréquence (Semblat & Pecker, 2009) :
- $Q^{-1}$ *est proportionnelle* à la fréquence pour le *modèle de Kelvin-Voigt* (ressort et amortisseur en parallèle),
- $Q^{-1}$ *est inversement proportionnelle* à la fréquence pour le *modèle de Maxwell* (ressort et amortisseur en série),
- l'atténuation peut présenter des variations plus complexes avec un *effet coupe-bande* pour le modèle de *Zener* (ressort en série avec une cellule de Kelvin).

Les résultats proposés sur la Figure 4 permettent de comparer les mesures expérimentales réalisées dans un massif de sol centrifugé avec des simulations issues des modèles de Maxwell et Kelvin. L'accélération de référence est le signal d'accélération délivré par le premier capteur (signal du haut). La coïncidence entre accélération mesurée (gauche) et simulations est assez satisfaisante pour le modèle de Kelvin (Figure 4, droite) et est excellente pour le modèle de Maxwell (Figure 4, centre).

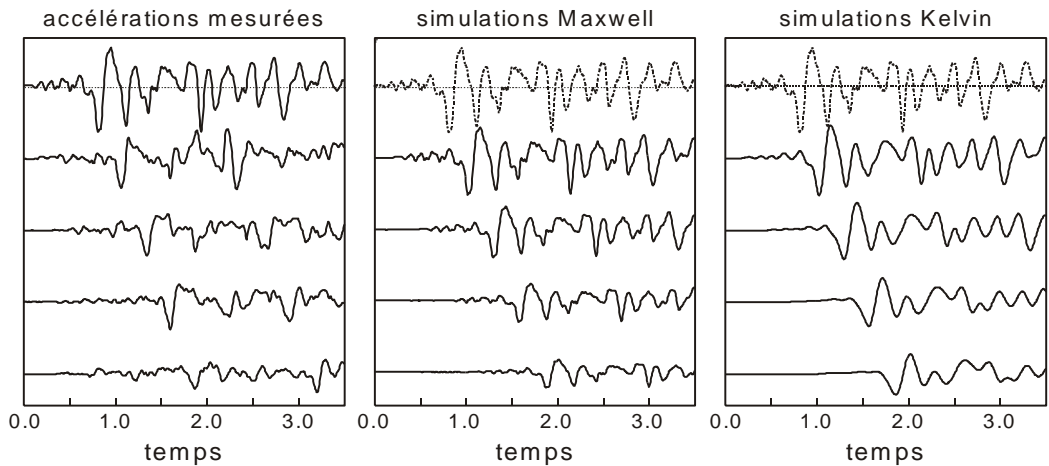

**Figure 4** : Signaux d'accélération mesurés en centrifugeuse (gauche) et simulations viscoélastiques à l'aide d'un modèle de Maxwell (centre) et d'un modèle de Kelvin (droite), d'après (Semblat et Luong, 1998).
*Acceleration signals measured in the centrifuge (left) and viscoelastic simulations considering a Maxwell (center) and Kelvin (right) model (Semblat & Luong, 1998).*

Les coefficients de viscosité des modèles rhéologiques utilisés dans les exemples de la Figure 4 sont les suivants : $\eta_{Kel}$=1000 Pa.s pour le modèle de Kelvin et $\eta_{Max}$=150000 Pa.s pour le modèle de Maxwell. Cela permet donc de caractériser l'amortissement dans le sol et d'utiliser ce résultat soit pour analyser finement les mesures obtenues sur site, soit afin de réaliser des simulations numériques.

### 4.2 Modélisation théorique de l'amplification

Dans le domaine sismique, il est prouvé qu'une onde peut être fortement amplifiée lorsque le contraste de célérité entre les couches de sol est important (Aki et Richards 1980, Bard et Bouchon 1985, Semblat et al. 2000, 2005). On rappelle donc ici comment l'amplification d'une onde de cisaillement dans une couche de sol peut être décrite dans le cas simple d'une couche homogène d'épaisseur constante (Figure 5, gauche). Il est possible de calculer les ondes se propageant dans la couche et dans le sol sous-jacent et d'estimer la fonction de transfert de l'onde à travers la couche.

Dans chaque milieu, le déplacement résultant de la superposition des ondes se propageant vers le haut ($z$>0) et vers le bas ($z$<0) s'écrit en choisissant l'origine de l'axe $z$ au sommet de chaque couche (Aki et Richards 1980, Semblat et Pecker, 2009) :

$$u_n = \left[ A_n \exp\left(ik_{z_n} z_n\right) + A'_n \exp\left(-ik_{z_n} z_n\right)\right] f_n(x,t) \tag{6}$$

où $k_{z_n}$ est le nombre d'onde vertical dans la couche *n* défini par :

$$k_{z_n} = \frac{\omega \cos \theta_n}{V_{S_n}} \tag{7}$$

$$f_n(x,t) = \exp\left[\frac{i\omega}{V_{S_n}}\left(x \sin \theta_n - V_{S_n} t\right)\right] \tag{8}$$

avec $V_{S_n}$ célérité des ondes de cisaillement et $\theta_n$ incidence dans la couche *n*.

L'indice *n* vaut 1 pour la couche de surface et 2 pour le sol sous-jacent. $A_n$ et $A'_n$ sont les amplitudes des ondes se propageant respectivement vers le haut et vers le bas dans la couche *n*. On peut déduire de l'analyse des différentes ondes la fonction de transfert à travers la couche.

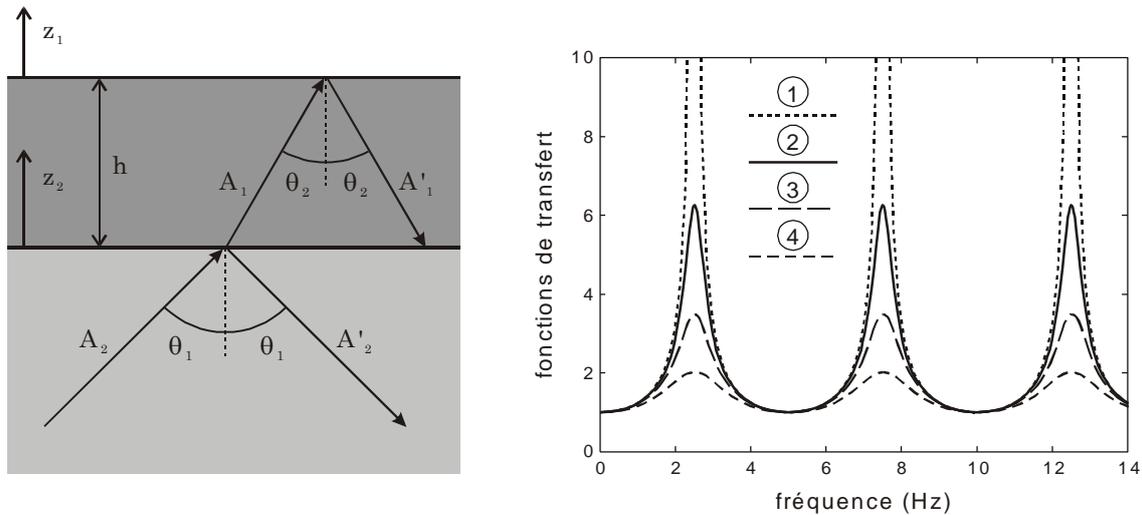

**Figure 5** : Amplification d'une onde plane dans une couche de sol (Semblat et Pecker, 2009) : définition des différentes ondes (gauche) et fonction de transfert pour différents rapports de célérités (droite).
*Amplification of a plane wave in a soil layer (Semblat & Pecker, 2009) : definition of the various waves (left) and transfer function for different velocity ratios (right).*

Par définition, la fonction de transfert entre deux points de la couche est le rapport des amplitudes de déplacement (ou vitesse, ou accélération) en ces deux points dans le domaine des fréquences. On peut alors définir la fonction de transfert à travers la couche de différentes manières (Semblat et Pecker, 2009) :
- *définition 1* : on considère le mouvement à la surface de la couche et le *mouvement à l'interface*,
- *définition 2* : on considère le mouvement à la surface de la couche et le mouvement en un point hypothétique situé à la surface du sol sous-jacent en l'absence de couche (*« mouvement affleurant »*). La 2$^{ème}$ définition coïncide avec la première lorsque le sol sous-jacent est supposé infiniment rigide.

La fonction de transfert issue de la 2ème définition s'exprime sous la forme:

$$\overline{T}(\omega) = \frac{1}{\cos k_{z_1} h + i\chi \sin k_{z_1} h} \qquad (9)$$

où :
$$\chi = \sqrt{\frac{\mu_1 \rho_1}{\mu_2 \rho_2}} \frac{\cos \theta_1}{\cos \theta_2} \qquad (10)$$

Cette fonction de transfert s'exprime donc en variables complexes et dépend des caractéristiques des deux milieux (ce qui n'est pas le cas avec la première définition). En considérant une couche de sol de célérité $V_{S1}$=200m/s, de masse volumique $\rho_1$=2000kg/m$^3$ et d'épaisseur 20 m, la Figure 5 (droite) compare plusieurs fonctions de transfert correspondant à des propriétés différentes dans la couche sous-jacente :
- cas 2 : $V_{S2}$=5000m/s, $\rho_2$=2500kg/m$^3$,
- cas 3 : $V_{S2}$=2000m/s, $\rho_2$=2200kg/m$^3$,
- cas 4 : $V_{S2}$=800m/s, $\rho_2$=2000kg/m$^3$.

L'influence du contraste de célérité dans les deux milieux apparaît clairement puisque, dans le cas 4, la valeur maximale de la fonction de transfert est inférieure à 2 alors que, dans le cas 2, le mouvement est amplifié d'un facteur 6 à certaines fréquences. Le cas 1 correspond à la définition 1 de la fonction de transfert équivalente à la définition 2 lorsque le sol sous-jacent est supposé infiniment rigide. En l'absence d'amortissement, l'amplification est dans ce cas infiniment grande à certaines fréquences (Figure 5, droite).

Pour les problèmes vibratoires, l'influence du contraste de célérité entre couches sera donc aussi très grande. Il conviendra toutefois d'utiliser une source adaptée (géométrie, directivité...).

### 4.3 Modélisation numérique de la propagation des vibrations

Pour analyser des cas réalistes, les modèles théoriques ne suffisent généralement pas. Les problèmes de propagation d'ondes sont en effet caractérisés par des phénomènes complexes (Eringen 1974, Semblat et Pecker 2009) : dispersion, diffraction, amortissement, conversions de type d'ondes... Il est souvent nécessaire de recourir à des simulations numériques et/ou des méthodes inverses (Bui 1993, Semblat et al. 2000) afin de déterminer les paramètres caractérisant le matériau et les ondes qui s'y propagent.

#### *4.3.1 Avantages et inconvénients des différentes méthodes numériques*

Plusieurs méthodes numériques permettent de simuler les phénomènes de propagation d'ondes : différences finies (Moczo et al. 2002, Virieux 1986), éléments finis (Joly 1982, Semblat 1998), éléments de frontière (Bonnet 1999, Chaillat et al 2009, Dangla 1989), éléments spectraux (Faccioli et al 1997, Komatitsch et al 1999). Suivant les applications visées, ces méthodes numériques présentent des avantages et des inconvénients différents.

La méthode des éléments finis (Figure 6, gauche) est très puissante car elle permet de modéliser des géométries et des comportements complexes (Ju et Lin 2004, Paolucci et al 2003). Pour les problèmes de propagation d'ondes, elle présente toutefois deux inconvénients principaux : la réflexion d'ondes parasites sur les frontières du domaine maillé, nécessitant des méthodes de frontières (ou couches) absorbantes (Chadwick et al. 1999, Collino 1996, Festa et Nielsen 2003, Meza-Fajardo et Papageorgiou (2008), Modaressi et Benzenati 1992, Semblat et al. 2010), et la dispersion numérique des ondes (Joly 1982, Hughes et al 2008, Ihlenburg et Babuska 1995, Semblat et Brioist 2000). La dispersion numérique provoque une variation artificielle de la vitesse de propagation des ondes en fonction des caractéristiques du modèle d'éléments finis (Figure 6, gauche).

La méthode des éléments de frontière (Figure 6, droite) présente l'avantage de permettre une modélisation aisée de la propagation d'ondes en milieu infini ou semi-infini (Dangla et al 2005, Chaillat et al 2009, Sheng et Jones 2006). Les conditions de radiation des ondes à l'infini sont en effet directement incluses dans la formulation. Par ailleurs, la méthode des éléments de frontière résout le problème aux interfaces entre milieux de caractéristiques homogènes : elle permet donc un gain sensible pour la modélisation de la propagation bidimensionnelle (interfaces unidimensionnelles) ou tridimensionnelle (interfaces surfaciques), mais est donc limitée à des milieux faiblement hétérogènes.

Afin de bénéficier des avantages de ces deux méthodes, il peut être intéressant de les combiner en réalisant un couplage éléments finis-éléments de frontière (Dangla 1989).

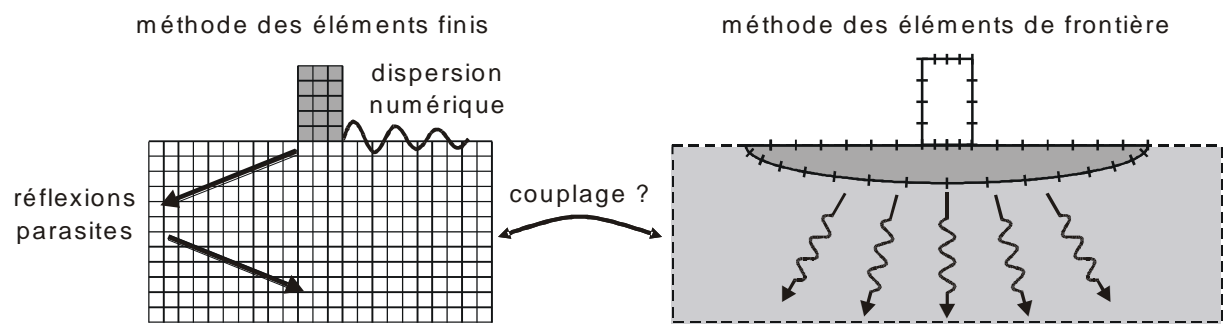

**Figure 6** : Méthode des éléments finis (gauche) et méthode des éléments de frontière (droite) pour modéliser les problèmes de propagation d'ondes et de vibrations.
*The Finite Element Method (left) and the Boundary Element Method (right) to model problems in the field of wave propagation and vibrations.*

*4.3.2 Exemple : modélisation des vibrations dans un tunnel ferroviaire*

L'exemple de simulation numérique considéré concerne un tunnel ferroviaire de 10 m de diamètre sous une couverture de 10 m (Figure 7). Le revêtement en béton a une épaisseur de 50 cm. Comme schématisé par une flèche sur la Figure 7, le tunnel est soumis à une sollicitation vibratoire due par exemple au passage d'un train (Semblat et Dangla 2005).

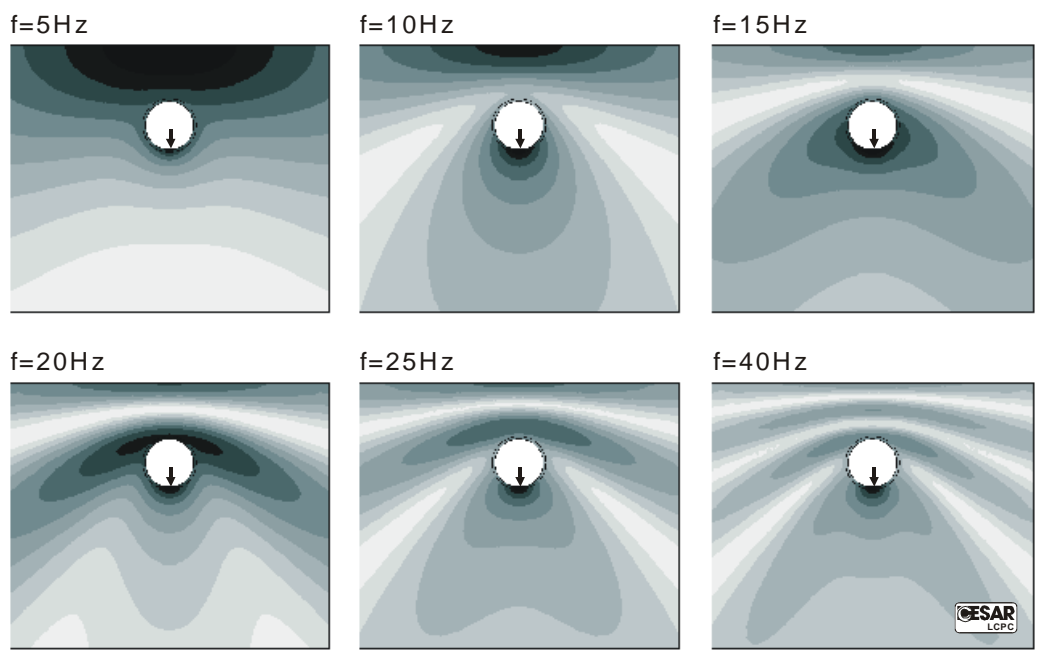

Figure 7 : Vibrations induites autour d'un tunnel par le passage d'un train : modélisation par la méthode des éléments de frontière (Semblat et Dangla 2005).
*Vibrations induced around a tunnel by the passage of a train: modelling by the Boundary Element Method (Semblat & Dangla 2005).*

La figure 7 montre les isovaleurs de déplacement calculées par la méthode des éléments de frontière à différentes fréquences. Les figures d'interférence, dues à la propagation de l'onde autour du tunnel, permettent de localiser les mouvements les plus forts du sol et notamment à l'aplomb de la charge ainsi qu'en surface.

Ce type de calcul a fait l'objet d'une comparaison avec les résultats d'une expérience de vibration en vraie grandeur sur un tunnel de la ligne Paris-Strasbourg (Dangla 1989). Les mouvements du sol ont été enregistrés à la surface et à la verticale du tunnel. La figure 8 montre les résultats de la comparaison calcul-expérience. Le modèle 2 se distingue du modèle 1 par la prise en compte de l'hétérogénéité du sol grâce à une modélisation couplant la méthode des éléments finis et celle des équations intégrales de frontière (Dangla 1989).

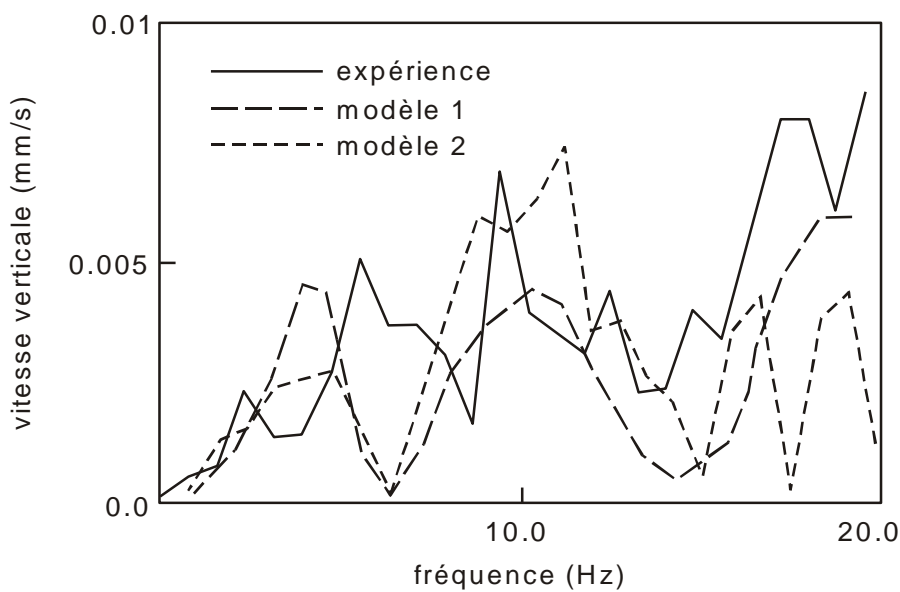

Figure 8 : Comparaison entre le calcul et l'expérience sur un tunnel réel
(Semblat et Dangla 2005).
*Comparison between the model and the experiment for an actual tunnel
(Semblat & Dangla 2005).*

Les calculs bidimensionnels ont été corrigés par un facteur théorique tenant compte de la géométrie hors plan du tunnel ce qui permet d'inclure de façon approchée l'effet tridimensionnel. Ces résultats sont dans l'ensemble assez satisfaisants puisque les amplitudes obtenues sont comparables. On doit tout de même observer un décalage en fréquence, au voisinage de 5 Hz, entre les courbes calculées et les résultats expérimentaux. De nombreux exemples de simulation sont proposés dans la littérature (Fiala et al 1987, Ju et Lin 1984, Paolucci et Spinelli 2006, Sheng et al 2006).

## 5  Isolation vibratoire

La compréhension des phénomènes liés à la propagation d'ondes issues de sources vibratoires peut donc résulter d'une analyse des données expérimentales de terrain, de laboratoire et de différentes simulations numériques. Cette connaissance facilite la conception de solutions nouvelles permettant de réduire l'effet des vibrations engendrées. Dans ce paragraphe son présentés un exemple historique d'isolation vibratoire (mesure sur le terrain et modélisation numérique), une expérimentation originale sur différents dispositifs d'isolation et pour différentes sources vibratoires et enfin des modélisations numériques récentes étudiant l'efficacité d'écrans enterrés.

### 5.1  Résultats de référence

Des expérimentations d'isolation vibratoire ont été réalisées dès les années 60 et 70 par Woods (1968) et Richart et al. (1970). Comme présenté sur la Figure 9 (en haut), ces auteurs ont analysé l'efficacité de différents dispositifs d'isolation vibratoire de type tranchées ouvertes (géometries circulaire, rectangulaire, etc). Une machine tournante a été installée à proximité de la tranchée afin de mesurer l'efficacité de celle-ci vis-à-vis de l'isolation vibratoire. Le facteur de réduction d'amplitude représenté sur la Figure 9 (en haut) est estimé en divisant l'amplitude de vibration mesurée avec la tranchée par l'amplitude de vibration obtenue sans la tranchée.

Leur analyse concerne donc la propagation d'ondes de surface (ondes de Rayleigh) créées par une source de vibrations harmoniques placée à la surface du sol pour des fréquences variant de 200 à 350 Hz. Woods (1968) a considéré différents types de tranchées caractérisées par leurs dimensions : profondeur *H*, largeur *w* et distance à la source de vibrations *R*. Ces paramètres géométriques sont donnés comme des fractions de la longueur d'onde de Rayleigh (ondes de surface), notée $\Lambda_R$ :
- Tranchée semi-circulaire : $H=0.60\Lambda_R$ et $R=0.60\Lambda_R$,
- Tranchée rectangulaire : $H=1.19\Lambda_R$ ; $R=2.97\Lambda_R$ et longueur $L=1.79\Lambda_R$.

Le facteur de réduction d'amplitude autour de la tranchée est représenté sur la Figure 9 (en haut) pour une tranchée semi-circulaire et une tranchée rectangulaire. Ces résultats montrent que des zones de faible amplitude (<0.125) apparaissent derrière la tranchée, alors qu'une amplification non négligeable (>1.25) peut survenir dans des zones situées en amont de la tranchée. L'efficacité du système d'isolation étudié par Woods (1968) et Richart et al. (1970) s'avère donc excellente.

Comme représenté sur la Figure 9 (en bas), des simulations numériques tridimensionnelles ont été réalisées sur ces deux configurations par Banerjee et al. (1988). Les résultats de ces simulations par la méthode des éléments de frontière prédisent le facteur de réduction d'amplitude de façon satisfaisante. La dissymétrie observée dans les résultats expérimentaux s'explique par le fait que les propriétés du sol réel ne sont pas homogènes.

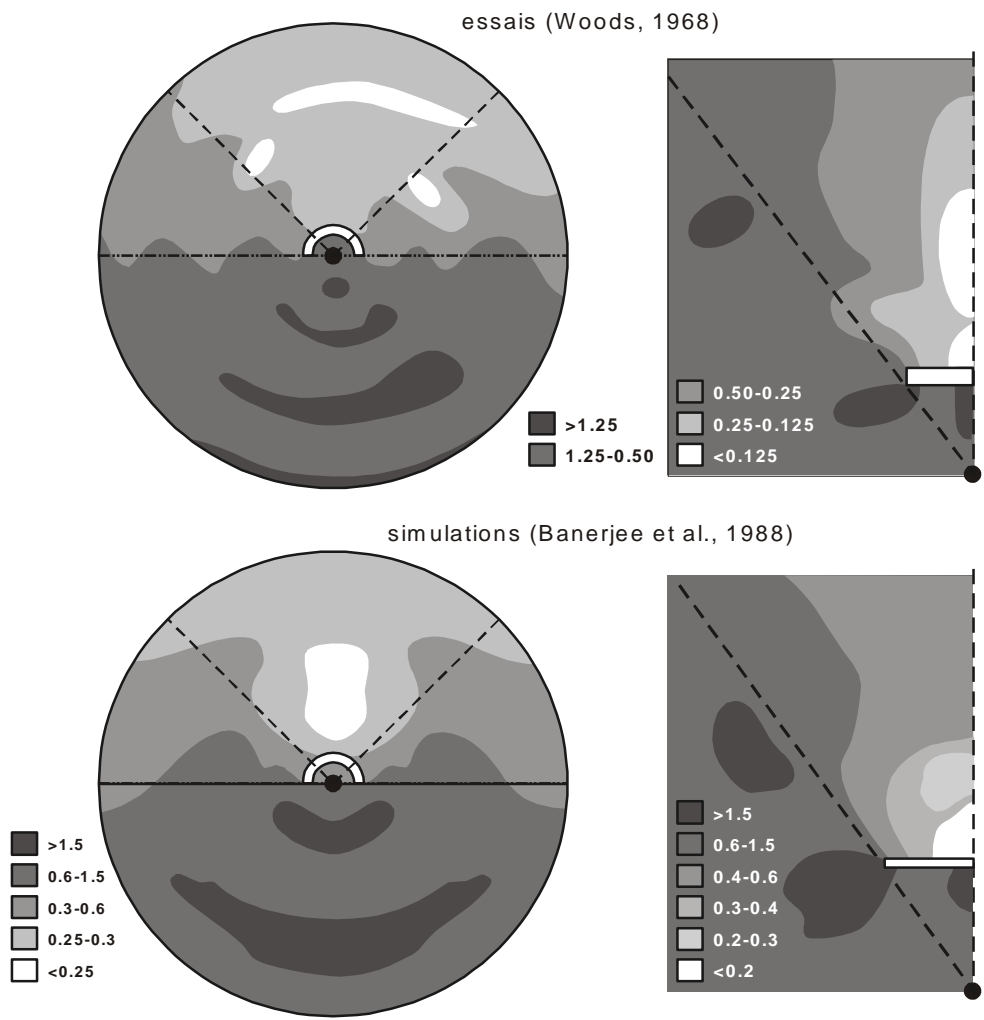

**Figure 9** : Isolation vibratoire à l'aide d'une tranchée circulaire (gauche) ou rectiligne (droite) : résultats expérimentaux de Woods (1968) et numériques de Banerjee et al. (1988), d'après Semblat et Pecker (2009).

*Vibration isolation using a circular (left) or linear (right) trench: experimental results from Woods (1968) and numerical ones from Banerjee et al. (1988), from Semblat & Pecker (2009).*

### 5.2 Analyse sur site contrôlé de différents systèmes d'isolation

L'efficacité de différents dispositifs d'isolation vibratoire a été récemment analysée sur site contrôlé par le Centre d'Expérimentation Routière (CER) de Rouen. Trois systèmes amortissants ont été considérés : chargement statique du sol, présence d'une paroi bétonnée de 5 m de profondeur, creusement d'une tranchée de 1, 2, 3 m de profondeur.

Ces essais ont été réalisés sur une structure de référence connue dite structure RCSU présentant 3 m de limon A1 surmontée de 2 m de grave C1B3 au sein du terrain naturel (grave argilo-sableuse avec affleurements de craie, classification des matériaux selon la norme NF P 11-300)) .

Les sources vibratoires utilisées sont : le portancemètre routier (source continue), la dynaplaque 2 et une chute de bloc béton de 1,25t dans certains cas (sources impulsionnelles). Les deux premières méthodes d'essais (portancemètre, dynaplaque) présentent l'avantage de créer une sollicitation dynamique analogue en intensité à celle provoquée par le passage d'un essieu à 60 km/h (4,5t/roue). Dans le présent article, seuls les résultats obtenus avec le portancemètre seront détaillés.

Les vibrations sont mesurées à différentes distances de la source au moyen de 5 à 6 géophones (Figure 10). Plusieurs modalités sont retenues :
- cas 1 : structure chargée à 8 endroits (contrainte minimale de 120 kPa) ;
- cas 2 : structure avec paroi en béton ;
- cas 3 : structure avec tranchée de 1, 2, 3m.

Trois capteurs de 2 Hz et trois capteurs de 1 Hz sont scellés au plâtre dans le terrain naturel à une hauteur correspondant au 2/3 de la hauteur du capteur. Les géophones 1Hz étant plus sensibles, ils ont été disposés en bout de ligne. Plusieurs campagnes de mesures ont été réalisées sur des structures de référence afin d'estimer l'amortissement apporté par les différents systèmes d'isolation.

| Système étudié | Mesure de référence |
|---|---|
| Chargement statique | Mesure sans chargement |
| Paroi bétonnée de 5m de profondeur | Mesure tranchée 0m |
| Tranchée 1, 2, 3m | Mesure tranchée 0m |

La différence mesurée entre les deux structures de référence (sans gueuse, sans tranchée) est liée à la portance du matériau au point d'impact (différence de 200MPa mesurée entre les deux structures). Les vitesses mesurées (Figure 11) donnent un amortissement sur la composante verticale de 78% avec les gueuses, de 74% avec la tranchée de 1 m et de 31% avec la paroi en béton. Ces essais indiquent que les tranchées de 2 et 3 m n'ont pas d'apport significatif sur l'amortissement, ce qui est lié à la configuration du dispositif (distance faible entre la source et le système amortissant). Suivant l'axe longitudinal, l'amortissement est de 57% avec les gueuses, de 11% avec la tranchée de 1 m et de 11% avec la paroi en béton.

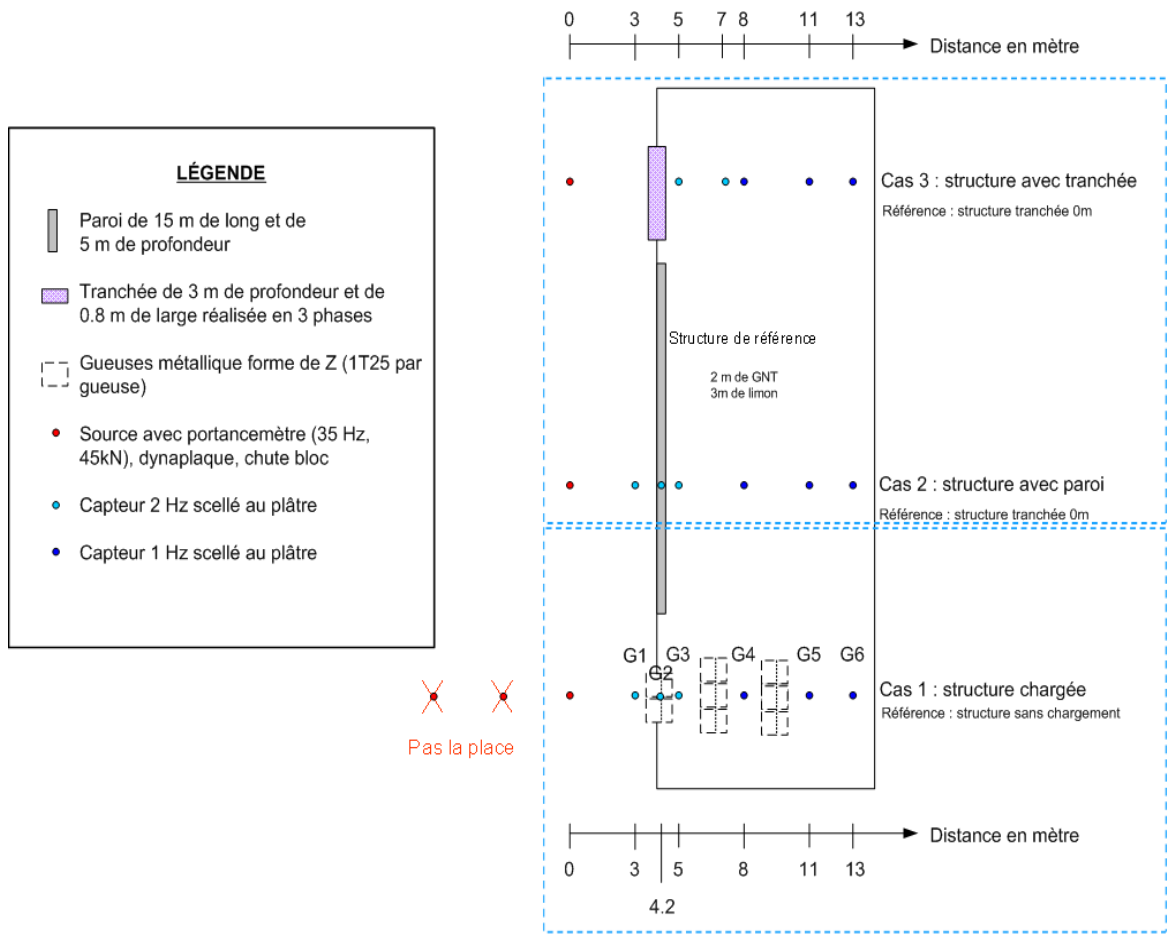

**Figure 10** : Implantation du dispositif expérimental pour évaluer trois systèmes amortissants (Jacqueline, 2009).
*Experimental setup to assess the three isolation systems (Jacqueline, 2009).*

Suivant l'axe transversal, l'amortissement est de 64% avec les gueuses, de 17% avec la tranchée de 1m et de 59% avec la paroi en béton.

Avec un écartement de 4,2 m entre la source et le système amortissant (tranchée, paroi bétonnée ou chargement statique), l'amortissement est maximal dans le cas de la structure chargée. La tranchée d'un mètre de profondeur amortit également bien la vibration suivant l'axe vertical mais ce dispositif présente un inconvénient d'un point de vue sécurité. L'apport des tranchées de 2 et 3m n'est pas visible pour le dispositif considéré ; la source étant proche de la tranchée, l'onde ne se propage pas en profondeur. Il conviendrait d'écarter la source du système amortissant pour voir l'effet de ces tranchées plus profondes.

D'après les résultats obtenus sur site contrôlé, l'efficacité du système d'isolation utilisant des masses en surface s'avère donc tout à fait intéressante. Suite à cette étude, il conviendrait d'étudier des barrières anti-vibratiles utilisant des matériaux amortissants (e.g. polystyrène), d'augmenter la distance source-dispositif pour estimer son influence sur l'efficacité des systèmes d'isolation et d'analyser l'effet de la fréquence de sollicitation sur l'amortissement dans les différents dispositifs.

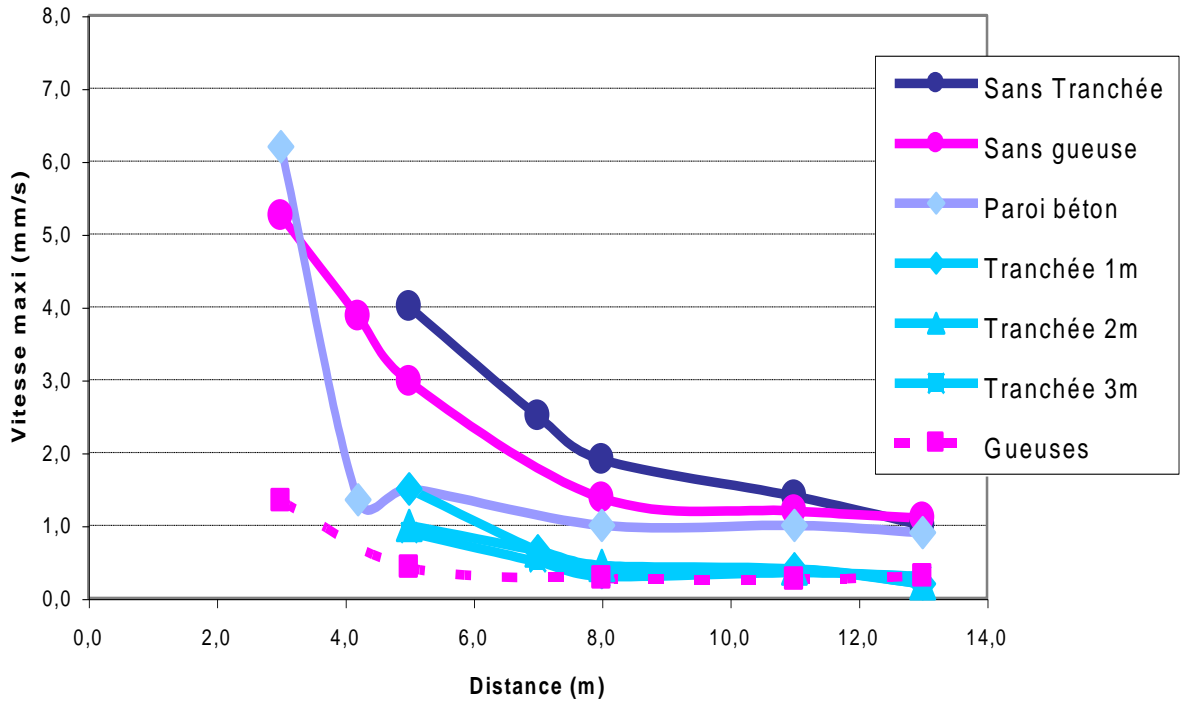

**Figure 11** : Comparaison des différents dispositifs atténuant les vibrations pour le portancemètre et la composante verticale.
*Comparison of various isolation systems for the vertical component (portancemeter loading).*

### 5.3 Modélisation de l'isolation

#### 5.3.1 Modélisation pour des configurations types

Dans le paragraphe précédent, l'efficacité de différents dispositifs antivibratoires a été analysée expérimentalement grâce à des essais sur site contrôlé. Un accord de recherche entre le LCPC, le CSTB et la RATP a par ailleurs permis de réaliser des tests numériques afin de quantifier la variabilité des résultats issus de différents codes de calcul aux éléments de frontière.

Dans le cadre de cet accord, la réponse de différents systèmes d'isolation a été analysée en considérant des sources ponctuelles et stationnaires sollicitant des fondations de formes variées. Une des configurations considérées est représentée en Figure 12. Les systèmes d'isolation sont constitués de parois enterrées en béton sans semelles ; une force verticale unitaire est appliquée au centre de la fondation en béton.

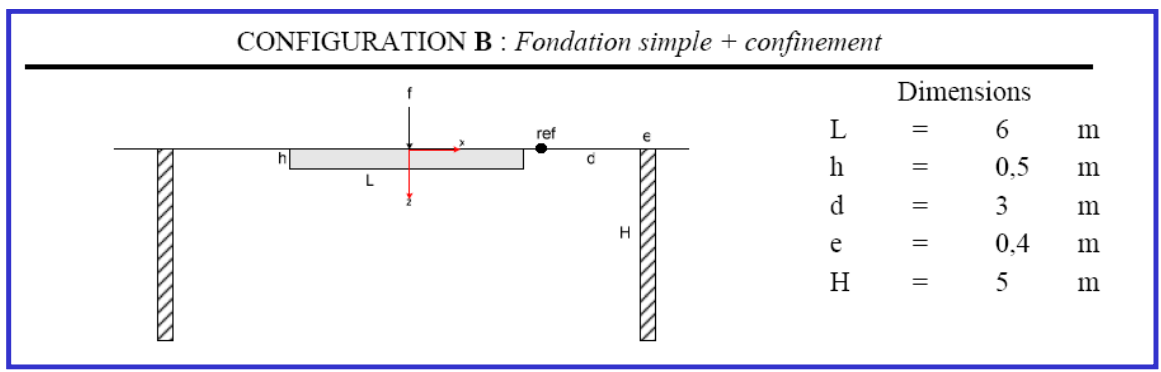

**Figure 12** : Configuration géométrique typique des cas-tests d'isolation vibratoire proposés par la RATP et le CSTB (Coquel 2008).

*Typical geometrical configuration to study vibration isolation
as proposed by RATP and CSTB (Coquel 2008).*

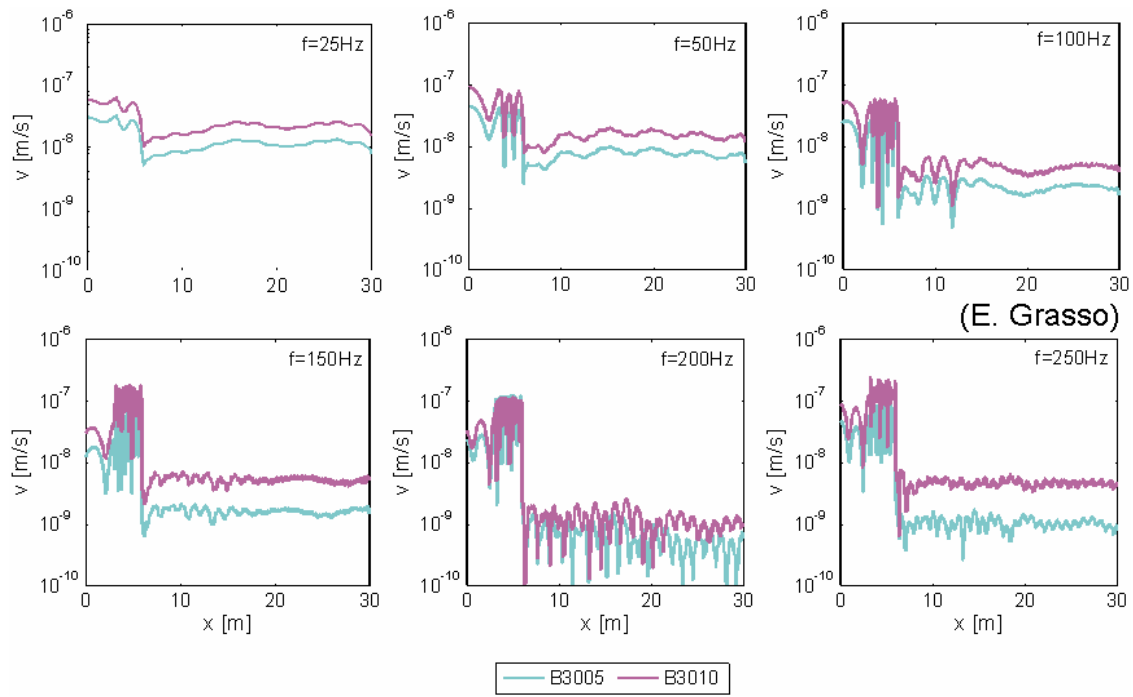

**Figure 13** : Résultats des simulations par la méthode des éléments de frontière du LCPC pour les cas-tests d'isolation vibratoire RATP/CSTB (Grasso 2008).
*Results of the BEM simulations performed at LCPC for the RATP/CSTB vibration isolation test-cases (Grasso 2008).*

L'utilisation de ce type de source permet d'apprécier la variabilité de l'atténuation en fonction de la fréquence. Sur la Figure 13, les résultats obtenus pour six fréquences (de 25 à 250 Hz) montrent les amplitudes en vitesse particulaire en fonction de la distance à la source. Elles mettent en évidence une diminution significative de l'amplitude au-delà de l'écran (axe y en échelle logarithmique). De plus, on remarque que l'efficacité de l'isolation dépend de la fréquence. En effet, à basse fréquence (e.g. 25 Hz), la longueur d'onde est grande comparativement aux dimensions de l'écran et l'efficacité de celui-ci est donc faible. En revanche, à haute fréquence (250 Hz), la longueur d'onde est nettement plus faible et l'efficacité de l'écran vis-à-vis de ces composantes spectrales est bien meilleure.

### 5.3.2  *Modélisation pour une configuration réaliste*

Karlström et Boström (2007) ont étudié l'efficacité de tranchées sur l'isolation vibratoire pour des configurations proches des problèmes ferroviaires. Deux cas sont comparés sur la Figure 14 : un cas sans dispositif d'isolation (en haut) et un cas avec une tranchée à gauche de la voie (en bas). Comme le montrent sur la Figure 14 les résultats numériques de Karlström et Boström (2007), l'efficacité de la tranchée est excellente puisque le niveau de vibration à gauche de la voie est fortement réduit. En revanche, les ondes mécaniques semblent piégées dans la structure de la voie puisque l'amplitude paraît sensiblement plus grande. Cela peut sans doute occasionner des dommages plus importants, ou des sollicitations en fatigue significatives, dans la structure de la voie.

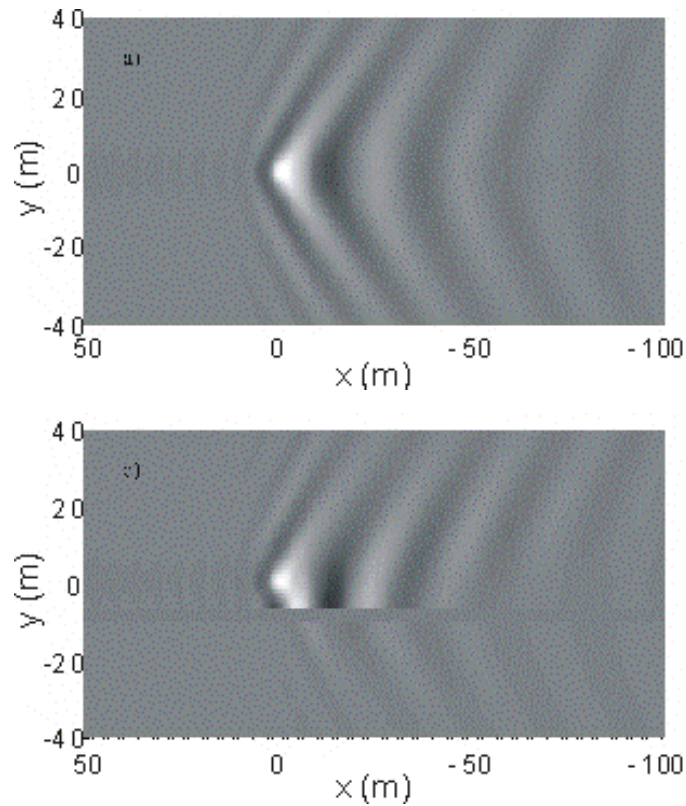

**Figure 14** : Modélisation d'un système d'isolation par tranchée dans le cas de vibrations ferroviaires (Karlström et Boström 2007) : pas d'isolation (en haut), une tranchée (en bas).
*Modeling of an isolation system with trench in the case of railway induced vibrations (Karlström & Boström 2007): no isolation (top), one trench (bottom).*

## 6  Conclusion

Les mesures de vibrations induites par le trafic ferroviaire permettent d'apprécier les vitesses particulaires maximales à proximité des voies mais conduisent généralement à des résultats complexes et difficilement généralisables. En outre, il peut s'avérer nécessaire de caractériser les sols et les sources vibratoires en laboratoire (e.g. colonne résonnante) ou en conditions contrôlées (e.g. fosse géotechnique, modèle réduit centrifugé) afin de d'accéder à certains paramètres dynamiques et d'appréhender les phénomènes prépondérants (atténuation, amplification, etc). Les essais en conditions contrôlées peuvent également permettre d'estimer l'efficacité de dispositifs d'isolation. Les résultats originaux proposés dans l'article montrent l'intérêt de dispositifs simples comme des masses posées en surface. Les paramètres identifiés sur site ou en laboratoire permettent ensuite de procéder à des simulations numériques pour des configurations simplifiées ou réalistes. Cela autorise une compréhension fine des phénomènes et des études paramétriques sur l'efficacité de différents dispositifs d'isolation vibratoire. Les cas tests analysés récemment avec la RATP et le CSTB ont ainsi permis de comparer numériquement différents systèmes d'isolation.

Enfin, pour aller au-delà, il serait également nécessaire d'étudier la transmission des vibrations aux bâtiments, la gêne occasionnée par les vibrations proprement dites ou par le bruit potentiellement émis dans les bâtiments (notion de « bruit solidien » (Coquel 2008, Fiala et al 2007)). L'évolution de la réglementation devrait tenir compte de l'ensemble des différents aspects.

## 7  Références